\documentclass{article}
\usepackage[centertags]{amstex}
\usepackage{amssymb}
\usepackage{epsfig}		

\makeatletter
\renewcommand\section{\@startsection{section}{1}{\z@}%
                                   {-3.5ex \@plus -1ex \@minus -.2ex}%
                                   {2.3ex \@plus.2ex}%
                                   {\reset@font\Large\scshape}}
\renewcommand\subsection{\@startsection{subsection}{2}{\z@}%
                                     {-3.25ex\@plus -1ex \@minus -.2ex}%
                                     {1.5ex \@plus .2ex}%
                                     {\reset@font\large\slshape}}

\makeatother

\def\np#1#2#3   {{\em Nucl.\ Phys.}\ #1 (#2) #3}
\def\pl#1#2#3   {{\em Phys.\ Lett.}\ #1 (#2) #3}
\def\prev#1#2#3 {{\em Phys.\ Rev.}\ #1 (#2) #3}
\def\prl#1#2#3  {{\em Phys.\ Rev.\ Lett.}\ #1 (#2) #3}

\newcommand\Tr{\operatorname{Tr}}
\newcommand\sgn{\operatorname{sgn}}

\begin{document}


\thispagestyle{empty}

\begin{flushright}
\large
SWAT/88 \\
DESY 96-054 \\
hep-lat/9604008
\end{flushright}

\vspace{1cm plus 1fil}

\begin{center}

\textbf{\huge Physical and unphysical effects in the mixed SU(2)/SO(3)
gauge theory}

\vspace{1cm plus 1fil}

{\large
{\bfseries P.W. Stephenson \\}
           Department of Physics, \\
           University of Wales, Swansea, \\
           Singleton Park, Swansea, SA2 8PP, U.K.\\
\vspace{0.1cm}
                       and \\
\vspace{0.1cm}
           DESY-IfH Zeuthen, 15735 Zeuthen, Germany\footnote{Present
           address.  Email: \texttt{pws@@ifh.de}}
}

\vspace{1cm plus 1fil}

\textbf{\Large Abstract}

\end{center}

\noindent 
We investigate possible problems with universality in lattice gauge
theory where a mixed fundamental SU(2) and SO(3)-invariant gauge group
is used: the (second order) finite temperature phase transition
becomes involved with first order effects with increased SO(3)
coupling, and this first order effect has a noticeable coupling
dependence for small lattices.  We produce evidence that the first
order transition is essentially bulk in nature as generally believed,
and that the finite temperature effects start to separate out from the
lower end of the bulk effects for a lattice of 8 sites in the finite
temperature direction. We strengthen our picture of the first order
effects as artefacts by using an improved action: this shifts the end
point of the first order line away from the fundamental SU(2) axis.

\newpage


\section{Introduction}
If the success of lattice gauge theory is to be measured by the
relevance of its results to experiment, we can now fairly claim that
a measure of success is finally arriving.  Our increased understanding
of the simulations and ability to improve them means that we can extract
interesting phenomenology.

This paper addresses a more basic worry in the subject: the
consistency of the continuum limit in the pure gauge theory.  The
choice of action is far from unique and one needs to be sure --- as a
prerequisite for the whole program of lattice gauge theory --- that
the action one chooses can reproduce the physical limit as the lattice
spacing $a$ is taken to zero.  Even when one decides (for simulational
convenience) on the Wilson form of the action, there is still an
ambiguity with regards to the representation of the gauge group.  This
matter was investigated and largely concluded more than a decade ago.

However, the subject was re-opened recently~\cite{ggm,mg}\ when it was
found that in one version of the SU(2) theory with couplings in mixed
representations the deconfining transition --- surely a truly physical
effect if that claim can ever be made for the pure gauge theory ---
was apparently mixed in with what had long been known as artefacts of
strong coupling.

This appears as a more fundamental problem than simply making the
continuum limit hard to obtain: if there is no clear separation
between the effects, one can never quite be sure that one's model
represents the physical theory sought.  In fact, two of the authors of
those papers have recently speculated~\cite{gm} that the nature
of the deconfining transition in SU(2), which was thought to have been
settled, may be under threat.

Since it is our intention in this paper to clarify what is, and what
is not, a physical effect which will survive the continuum limit, we
spend a certain amount of time in the next section explaining the
previous results concerning the bulk transitions (an introduction to
the early work on the subject is given in reference~\cite{creutz}).
We emphasise that, while these effects are not physical as far as the
underlying gauge theory is concerned, they are nevertheless not a
hindrance to a well-defined continuum limit.  We then introduce the
new problem with the deconfining transition.

In the following section, we present results from new simulations in a
(similar but not identical) variant of the theory aimed at clarifying
the position.  Finally, we attempt to draw all the results together
and suggest there is no danger to the continuum limit.

We should note straight away that we are using the word `artefact' to
denote anything obscuring the physics of the continuum limit of the
gauge theory in which one is interested; we are not necessarily
claiming that the other effects are uninteresting in their own right.
We use the words `physical' and `physically' with similar thoughts in
mind.  Further, we recognise that any such artefacts are a sign that
we are still far from the continuum limit; here we must inevitably
plead the excuse of finite computing resources.

A brief outline of early results has appeared in \cite{me}; our
investigations here are more detailed and our conclusion is different.


\section{The mixed action theory}

The Wilson action is very widely used in lattice gauge theory
as the basis for simulations because of the elegant and simply way it
retains gauge invariance in the discrete theory:
\begin{equation}
  S_W \equiv - \beta_\mathrm{rep}\sum_\Box \Tr_\mathrm{rep}U(\Box)
\end{equation}
in which $\beta_\mathrm{rep}$ is proportional to the reciprocal of the
bare coupling squared and the sum is over all plaquettes (closed loops
of the smallest possible size:  we use the symbol $\Box$ throughout to
represent a plaquette) on the lattice.  The variable $U(\Box)$ is the
element of the gauge group corresponding to this path.

The point here is that one must make a choice of the representation in
which one takes the trace of the group-valued $U(\Box)$.  The
fundamental representation is the most natural, as it corresponds to
the matrices with which one actually implements the theory
computationally and is also the representation usually associated with
fermions.  (There is then a conventional factor of $1/N$ in the
trace for SU(N).)

However, in the continuum the theory does not involve elements of the
gauge group at all: it is defined in terms of a Lie algebra.  The
gauge elements were introduced as a useful way of keeping track of
gauge invariance, reducing it to nothing more than the linear matrix
algebra which is so natural for a computer.  Therefore, in the pure
gauge theory at least, one should not be restricted to the fundamental
representation; one should obtain the same continuum theory from any
representation as the cut-off is removed.  We shall refer to this in
the current paper as \textit{universality}.  In general, the question
of universality refers much more widely to independence of the
continuum limit from the form of the discretisation, which depends on
a great many details; here we concern ourselves only with the
particular restricted form.  We must also recognise that on any finite
lattice, two distinct formulations are highly unlikely to be
identical.  Nonetheless, as we shall describe, there is a clear sense
in which the theories we are dealing with have the same continuum
limit.

Let us now specialise to the pure gauge theory with gauge group SU(2),
in which the fundamental representation is `spin-$1/2$', to use the
familiar language.  The group theoretical argument suggests that we
could just as well take the adjoint (spin-1) representation and still
see the same physics.  This representation does not show the
full SU(2) invariance.  The gauge manifold is a three-sphere $S^3$;
the adjoint representation is insensitive to factors of $\pm1$
(corresponding to the center Z(2) of the gauge group), so that
opposite ends of diameters on $S^3$ should be identified.  The actual
invariance shown is that of the gauge group SO(3): more generally,
this is true of all the whole-integer spin representations of SU(2),
while the half-odd-integer representations faithfully reproduce the
full invariance.  (We shall use the somewhat loose terminology
`SO(3)-' and `full SU(2)-invariant' where appropriate to distinguish
these cases.)  This matter of differing topology will turn out to be
crucial.

The first surprise~\cite{hs1} was that the SO(3)-invariant theory
turned out to have a strong first order transition not seen in the
full SU(2) theory.  We return to this below.

It was noticed by Bhanot and Creutz~\cite{bc}\ that one could combine
different representations of SU(2) (say, the fundamental and adjoint)
linearly into the same action, producing a two-dimensional parameter
space which would enhance one's ability to test universality:
\begin{equation}
  S_M \equiv - \frac{\beta_F}{2} \sum_\Box \Tr_F U(\Box)
  - \frac{\beta_A}{3} \sum_\Box \Tr_A U(\Box) \label{eqn:fundadj}
\end{equation}
with obvious notation and the conventional normalising
factors.  The SO(3)-invariance of the adjoint
term manifests itself as a squaring of the trace of the matrix
representing $U(\Box)$; $\Tr_A \sim (2\Tr_F)^2-1$.  Naively, one
would define a combined coupling $\beta_\mathrm{eff} =
\beta_F/4+2\beta_A/3$:  then in the regime where lattice effects were
negligible a physical theory would arise depending only on this
effective coupling.


\subsection{Bulk transitions}

\begin{figure}[tb]
  \begin{center}
    \psfig{file=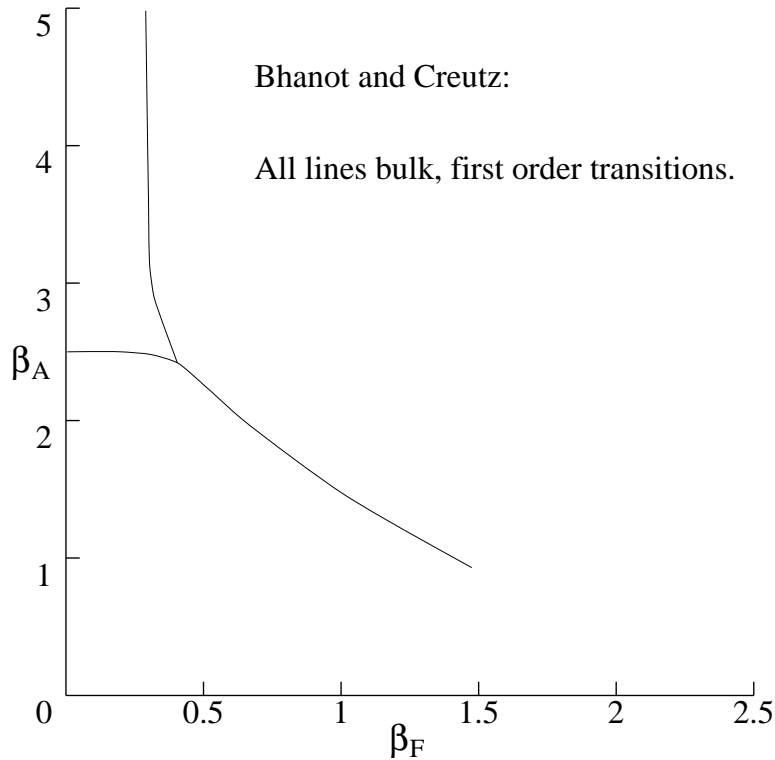,width=4in}
  \end{center}
  \caption{First order, bulk phase transitions found in the mixed
    fundamental-adjoint SU(2) theory by Bhanot and Creutz~\protect\cite{bc}.}
  \label{fig:bc}
\end{figure}
The resulting phase diagram showed that the SO(3) transition extended
into the plane from the $\beta_F=0$ axis, and combined with another
roughly vertical transition; the combined line ended abruptly in the
middle of the plane.  This is shown in figure~\ref{fig:bc}.  (To jump
ahead, our conclusion will later be that this is a complete picture of
the phase transition artefacts afflicting the gauge theory, so this
can be compared with our summary diagram, figure~\ref{fig:newres}.)

It appeared that the transitions were all of a bulk nature, in other
words independent of the size of the lattice (although it should be
noted that due to the computational limitations of the time the
simulations were restricted to $N_S^3\times N_T = 4^4$ lattices).
Thus, the transitions have no physical scale associated with them; as
the lattice spacing is taken to zero, and hence from asymptotic
freedom the inverse bare coupling $\beta$ is taken to infinity the
transitions remain behind at the same fixed coupling.  This means that
--- provided a continuum theory exists at all --- they are artefacts
of some sort.

There was one clue to the nature of the near-vertical part of the
transition: if $\beta_A$ is taken to infinity keeping $\beta_F$
finite, the fields are forced to $\pm I$ ($I$ is the group identity
element), which the adjoint representation does not distinguish.  Thus
there is an embedded Z(2) theory corresponding to flipping the sign of
any element of the gauge group; this becomes an exact symmetry for
zero $\beta_F$ at all $\beta_A$.  In the case of $\beta_A\to\infty$
for finite $\beta_F$, the coupling is just that of a Z(2) gauge model,
and the end of the line at $\beta_A=\infty$ is just the $Z(2)$
symmetry breaking transition of this model.  The coupling in this
limit is known analytically to be $\log(1+\sqrt{2})/2\sim0.4407$, which
agrees with the line appearing in the mixed action diagram.  Clearly,
in this limit the gauge theory no longer plays a part.  This leads us
to label the line in the diagram as an artefact.


\subsection{The Villain form}
Further work by Caneschi, Halliday, Schwimmer~\cite{hs2,chs} clarified
the nature of the SO(3)-like bulk transition.  They defined a
`Villain' form for the theory, in which the SO(3) invariant part of
the action was rewritten to include an auxiliary Z(2)-valued field
$\sigma(\Box)$, taking values on plaquettes.  The action becomes
\begin{equation}
S_V \equiv - \frac{\beta_F}{2} \sum_\Box \Tr_F U(\Box) -
\frac{\beta_V}{2} \sum_\Box 
\Tr_F U(\Box)\times\sigma(\Box) \label{eqn:villain}
\end{equation}
and the measure is extended to include a sum over ${\sigma(\Box)} =
{\pm1}$.  The SO(3) invariance is now manifest in this new $Z(2)$
symmetry: for every configuration with a given $U(\Box)=1$, there is
another with $U(\Box)=-1$.

If one is probing the differences between the full-SU(2) and SO(3)
theories, this form of the action is as good a tool as the
fundamental/adjoint form.  However, the Villain part does not
correspond to an irreducible representation of SU(2): in fact it
includes contributions from all representations with integer
spin; the expansion is given in reference~\cite{hs2}.  The naive
effective coupling in this case is simply
\begin{equation}
\beta_\mathrm{eff} = \beta_F + \beta_V.
\label{eqn:beff}
\end{equation}

The phase diagram found is very similar to the fundamental/adjoint
one, although the vertical axis has a different scale with the SO(3)
phase transition now around $\beta_V\sim4.3$ instead of $\beta_A=2.5$
(in fact, this is the only substantial difference between the theories
that we have noticed).


\subsection{Monopoles and charges}
The Villain form has both practical advantages (discussed in
the next section) and theoretical ones, namely the transparency with
which Z(2) effects can be seen in the behaviour of the $\sigma(\Box)$
variables.  The behaviour of the SO(3) transition was elucidated in
terms of the Z(2) effects in reference~\cite{hs2} and this was extended to
the mixed-action plane in reference~\cite{chs}; this is an expanded and
slightly re-interpreted account of the explanations therein.

The Z(2) degrees of freedom can be divided up into two types of
object, `monopoles' $M$ and `charges' $E$, defined by
\begin{equation}
  \begin{split}
    M(c) &\equiv \frac{1}{2}\left(1 - \prod_{\Box\in\partial c}
    \sigma(\Box)\right) \\
    E(l) &\equiv \frac{1}{2}\left(1 - \prod_{\Box\in\hat\partial l}
    \sigma(\Box)\right) \\
  \end{split}
  \label{eqn:monopole}
\end{equation}
in which $c$ is a (three-dimensional) cube and $l$ is a link of the
lattice: the monopole is defined as a product of the plaquette-valued
Z(2) variable over the faces of the cube $c$ and the charge as the
product over all plaquettes having the link $l$ in their perimeter.
Each can take the value $1$ or $0$.

In this picture, a cube having $M(c) = 1$ contains a monopole; one can
draw a Dirac string from it to another cube having $M(c)=1$ by
tracing plaquettes with $\sigma(\Box)=-1$.  

Considering the special case $\beta_F=0$, the charge degrees of
freedom are trivial: multiplying $E(l)$ by $-1$ is the same as
multiplying the link $U(l)$ by $-1$, so that the gauge variables and
the monopoles are sufficient to describe the complete theory.  One can
alter the values of the charges by flipping the signs of all relevant
plaquettes without changing the physical state.  As each cube contains
exactly two (or zero) plaquettes from a charge $E(l)$, this does not
change the value of a monopole either: this corresponds to an
unphysical movement of the Dirac string.  The authors of
reference~\cite{chs} suggest some dual behaviour between the monopoles
and charges in the mixed-action theory.

As the monopole is a Z(2) object, in the gauge theory one can think of
it as flipping the sign of a gauge element.  This is related to the
disconnected nature of the SO(3) gauge manifold; one can transform a
gauge element continuously from a value $g$ to $-g$, but as these
points are identified this is a closed path which cannot be shrunk to
zero.  The monopoles are a sign that such closed paths are
contributing to the path integral.

Thinking of monopoles as dynamical degrees of freedom, then, at small
$\beta_V$ entropy effects dominate and monopoles are present.  As
$\beta_V$ increases entropy loses out to minimising the action and the
gauge degrees of freedom tend to settle close to the identity.  In
this second case, the closed paths joining the regions around the
identity and its negative are not present, since gauge elements
representing plaquettes which lie in between would produce a large
action.  Hence the regions around the identity and its negative,
although images of one another in the pure SO(3) case, appear
disconnected when considering the whole three-sphere of the SU(2)
manifold: in other words, monopoles are suppressed.  The part of the
first order transition which survives in the limit $\beta_F\to0$
corresponds to the disappearance of the monopoles.

If monopoles are suppressed in the charge-independent pure SO(3) theory,
then in the second term of equation~\ref{eqn:villain} we are
left with only $\beta_V\Tr_F U(\Box)$, identical to the fundamental
theory.  It was indeed found by Halliday and Schwimmer~\cite{hs2} that
the monopoles were strongly suppressed in the high-$\beta_V$ phase;
thus the continuum limit of the SO(3) theory is expected to be the
same as that of SU(2) once the phase transition is passed (though the
approach to the continuum may be different due to residual monopole
effects).

Including the other piece of information about the bulk transitions,
namely the Z(2) gauge model limit, the nature of the boxed-in corner of the
mixed-action phase diagram becomes clearer.  The Z(2) symmetry which
was manifest for the charges in the SO(3) theory survives with
increasing $\beta_F$ out to the first order phase transition and is then
broken.


\subsection{Understanding the first order effects}

Here we summarise what the monopole/charge picture tells us about the
bulk transitions.  It is to be remembered that we are everywhere
talking about the bare degrees of freedom, i.e. those defined directly
on the lattice, rather than the physical fields for which the picture
can be very different.

One can distinguish the upper left corner of the mixed action
diagram from the rest of the plane by the following: there, the
underlying gauge system occurs around the identity $I$ of SU(2) as
well as an image around $-I$.  In this region there are no gauge
elements lying near the `equator' of the gauge manifold because the
action for that is too great, so topologically non-trivial closed
paths are not important.  The two systems around $I$ and $-I$ are
related by an exact Z(2) symmetry for $\beta_F=0$; the Monte Carlo
results show that the effect of this symmetry persists to finite
$\beta_F$.

In the rest of the mixed-action plane, there is only one gauge system
rather than the two images.  For increasing $\beta_F$, this is simply
the usual theory localised more and more around the identity.  In the
special case $\beta_F=0$ for $\beta_V$ below the monopole transition,
there is still only one gauge system, but the fields are spread over
the whole manifold of the group: increasing $\beta_F$ causes a smooth
breaking of the Z(2) symmetry.

Thus, in whatever direction we choose to take the continuum limit, we
have a smooth transition to the perturbative regime either around $\pm
I$ or $I$ alone.  The only exception is the limit $\beta_V\to\infty$
with $\beta_F$ finite.  This is pathological because one is
effectively tuning away the gauge system leaving only the Z(2)
variables; in every other direction it is the Z(2) degrees of freedom
which become irrelevant, either due to suppression of monopoles, or to
breaking of Z(2).  Hence the conclusion is that universality is not in
danger from these bulk effects.


\subsection{Finite temperature effects}
Recently, however, this simple picture was confused by new results
in the region of the tail of the bulk transitions, where they join
together and apparently reach an end point.  Gavai, Grady and
Mathur~\cite{ggm,mg} followed the finite-temperature transition,
well-known in fundamental SU(2), into the fundamental adjoint plane.
It is to be emphasised that this transition is physical, having been
comprehensively investigated~\cite{efmwhk1,efmwhk2,efmwhk3} and shown
to obey scaling.  With the critical temperature on an $L_S^3\times L_T
\equiv (N_Sa)^3\times(N_Ta)$ lattice being $T_c=1/L_T=1/(N_Ta)$, the
transition moves to smaller $a$ and hence larger $\beta_F$ as the
number of lattice sites in the time direction $N_T$ is increased.
Thus we would not expected it to be involved with the bulk effects
occurring at fixed coupling.

(Even this naive picture presumably has to be modified in some way as
one reaches the SO(3)-invariant axis, since confinement dynamics is
different due to the lack of anything like a string of fundamental
flux~\cite{dgh}.  Nonetheless one clearly does not expect bulk effects
to be involved.)

However, it was found that on the contrary the phase transition's
extension for finite $\beta_A$ pointed directly towards the tail of
the bulk transition, and indeed for $N_T=4$ turned into the transition
which Bhanot and Creutz on their $N_T=N_S=4$ lattices had thought to
be bulk.  The transition changed from second to first order; there was
no evidence for separate bulk and finite temperature effects at any
$\beta_A$.

The problem, therefore, is to find some way of separating the
artefacts (the bulk transitions we thought we understood) from
the physics (here, the finite temperature transition).  This is the
problem we address in the remainder of the paper.



\section{New simulations}

We have performed simulations using the Villain form of
equation~\ref{eqn:villain}.  This has the advantage over the
fundamental/adjoint form (equation~\ref{eqn:fundadj}) that it is linear in
the matrices actually used in the simulation.  One is able to perform
the Monte Carlo update in two parts.  First, the gauge fields are
updated; the extra $\sigma(\Box)$ parts are here treated as a
modification of the `staples' multiplying the central link at each
stage of the update.  Thus a standard heatbath approach can be used:
we have used the form due to Kennedy and Pendleton~\cite{kp}, though
we have not made any detailed evaluation of its performance in the
Villain theory.  Next, the Z(2) variables are updated with the gauge
variables constant; this can again be done by a standard heatbath and
is particularly simple as the Z(2) variables are not directly coupled
to one another.  One can also apply exact overrelaxation to the gauge
fields.  This is in contrast to the adjoint case where one is limited
to a less efficient $N$-hit Metropolis update.  In general, we have
adopted the fairly standard procedure of using four overrelaxation
steps of the entire lattice to every heatbath step.  In what follows,
this compound step is referred to as a single sweep.

Calculations were performed on every sweep.  We calculate the action
(fundamental and adjoint), the Polyakov loop in the time direction and
one spatial direction, and the Halliday-Schwimmer monopole and charge
values as well as the effective monopole and charge values obtained by
using the sign of the plaquette instead of the Z(2) variable itself,
found by replacing $\sigma(\Box)$ by $\sgn \Tr_F U(\Box)$ in
equation~\ref{eqn:monopole}:
\begin{equation}
  \begin{split}
    \bar M(c) &\equiv 1 - \prod_{\Box\in\partial c} \sgn \Tr_F U(\Box) \\
    \bar E(l) &\equiv 1 - \prod_{\Box\in\hat\partial l} \sgn \Tr_F U(\Box). \\
  \end{split}
  \label{eqn:effmonopole}
\end{equation}
The temporal Polaykov loop is defined by
\begin{equation}
  P_t \equiv {\sum_{x,y,z} \frac{1}{2} \Tr_F \prod_{t=0}^{N_T}U_t(x,y,z,t)
    \over N_S^3}
\end{equation}
and similarly for the spatial value $P_x$.

In the case where the transition appears to be the well-known second
order one, our main interest is in the order parameter $P_t$, whose
symmetry breaking signals deconfinement.Given the symmetry breaking,
we are then interested in locating a peak in the susceptibility of the
absolute value of the Polyakov loop,
\begin{equation}
  \chi_{P_t} \equiv N_S^3(\langle P_t^2\rangle
  - \langle\lvert P_t\rvert\rangle^2)
\end{equation}
which we interpret as the position of the phase transition.  Strictly,
there can be no phase transition of this nature on a finite lattice;
this is one standard and convenient procedure which we adopt here.
(We also adopt the useful fiction of referring to the crossovers as
phase transitions where we believe they would become so on infinite
lattices.)

We can also use $\chi_{P_t}$ to help us identify the order of the
phase transition:  for lattices $N_S^{(1)}$ and $N_S^{(2)}$ with the
same $N_T$ we form the ratio
\begin{equation}
  \begin{split}
    R(N_S^{(1)},N_S^{(2)}) & = \frac{{N_S^{(1)}}^3*\chi_{P_t}(N_S^{(1)})}
    {{N_S^{(2)}}^3*\chi_{P_t}(N_S^{(2)})} \\
    & \equiv \left(\frac{N_S^{(1)}}{N_S^{(2)}}\right)^\omega \\
  \end{split}
  \label{eqn:ratio}
\end{equation}
For a first order transition, the effect behaves like the volume of
the lattice, so that the exponent $\omega=3$.
For a second order transition in the same universality class as the
the three-dimensional Ising model --- as the usual finite temperature
transition in fundamental SU(2) appears to be --- the value is
$\omega=1.97$.  (See references~\cite{efmwhk1,efmwhk2,efmwhk3} for
more detailed discussions of the SU(2) finite temperature transition.)

We have used the density of states method (also known as
Ferrenberg-Swendsen reweighting~\cite{fs}) to locate the peak and when
located to trace its outline; again, our procedure is entirely
standard.

We have started our exploration using lattices with time sizes $N_T=2$
and $N_T=4$, extending this to larger lattices where this seems
warranted.  It should be made clear that it is not our goal explicitly
to extract continuum physics from the systems under consideration; in
fact, we shall maintain that this is in practice impossible in many
cases.  The goal here is to understand qualitatively the rather
puzzling features seen in the theories.  Hence our lattices are simply
chosen to be as large as we need to identify trends in the data and
are not designed, for example, for a full scaling analysis.

Instead of listing results piecemeal as we proceed, all the results
are given in tables~\ref{tab:ptpos} to~\ref{tab:ptmeta}, to which we
shall refer back.  The tables are divided such that
tables~\ref{tab:ptpos} (position of phase transitions)
and~\ref{tab:ptsus} (the susceptibility $\chi_{P_t}$) contain results
where we have performed long runs (typically $200,000$ sweeps) and
used reweighting to determine the position of and height of the peak
in the susceptibility $\chi_{P_t}$.  Table~\ref{tab:ptmeta} contains
all the remaining runs, where this procedure is impossible because of
the metastable effects with a long time to flip between states and we
have merely located the position of the transition by the methods
described.  Note that while it seems clear that all entries in
table~\ref{tab:ptmeta} describe first order effects --- at any rate
something incompatible with the usual second order finite temperature
behaviour --- it is not necessarily safe to conclude that those in
tables~\ref{tab:ptpos} and~\ref{tab:ptsus} are necessarily second
order; a weak first order signal is notoriously difficult to
disentangle from this.

\begin{figure}[tb]
  \begin{center}
    \psfig{file=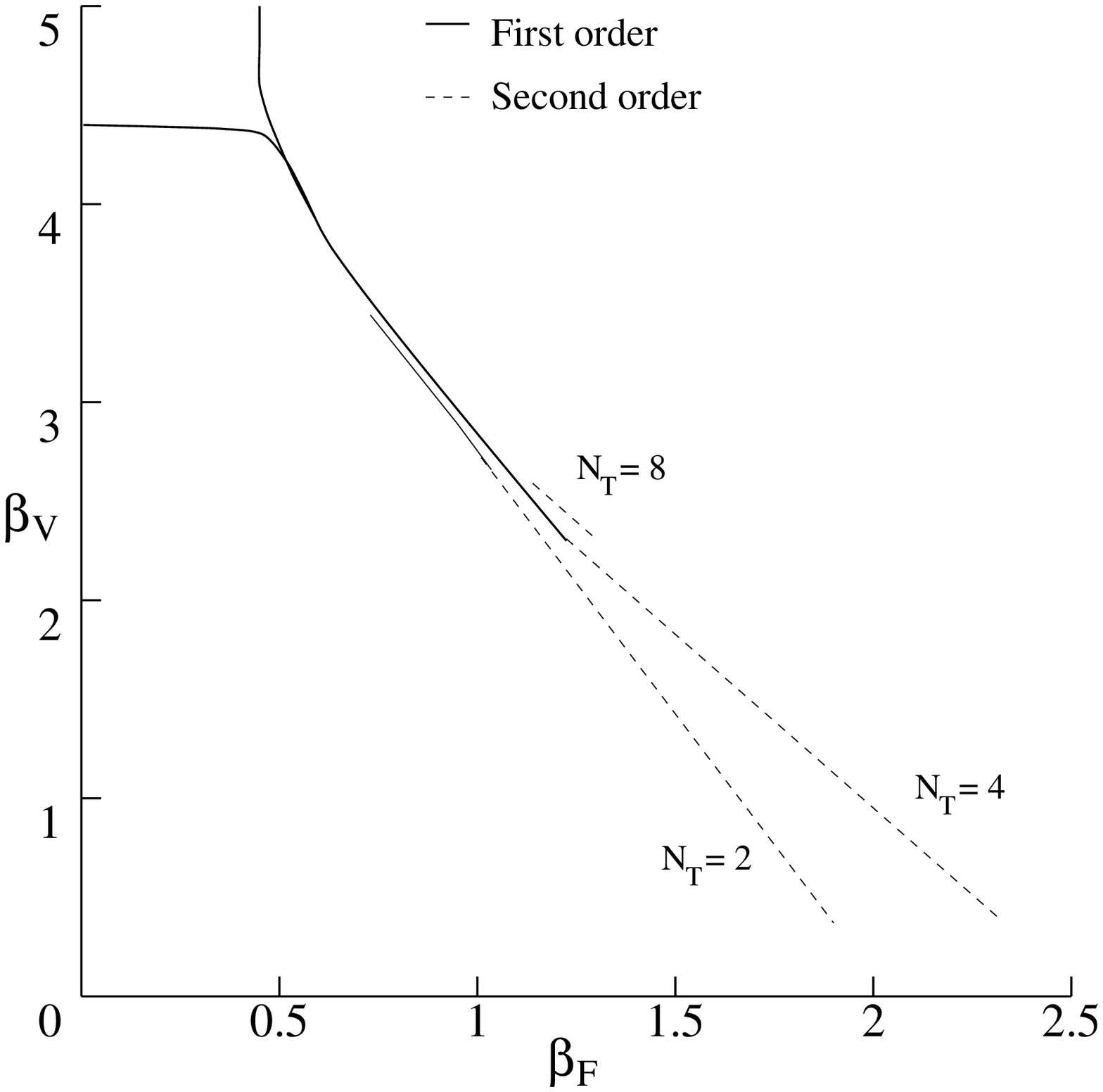,width=4in}
  \end{center}
  \caption{Our picture of the fundamental/Villain plane, sketching the
    phase transitions for three values of $N_T$.  We interpret the
    shifted first order line of $N_T=2$ as an additional effect due to
    small lattice size.  Note that for $N_T=8$ for $\beta_V=2.4$ we
    see both the first order and second order lines.}
  \label{fig:newres}
\end{figure}

\begin{figure}[tb]
  \begin{center}
    \psfig{file=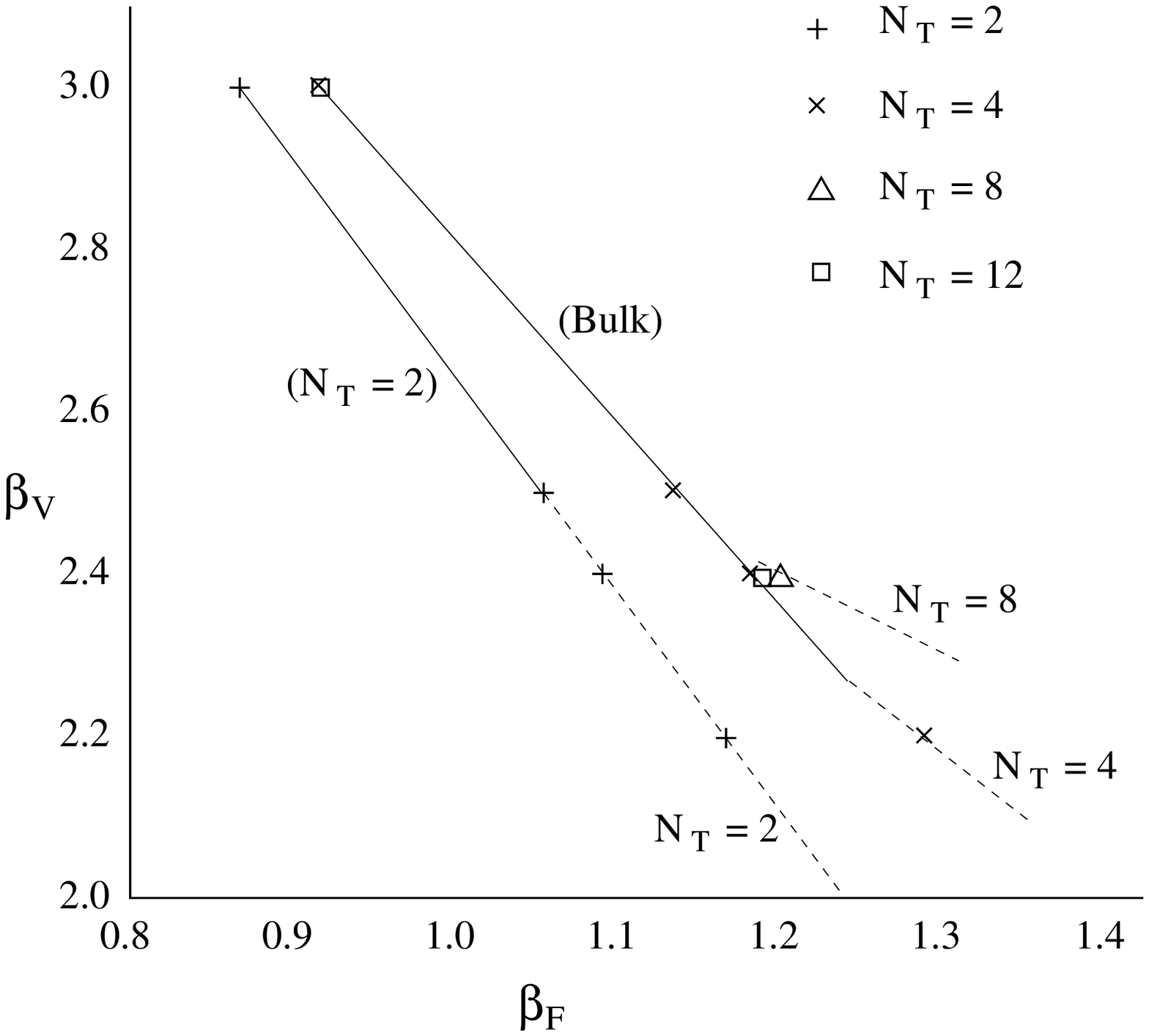,width=4in}
  \end{center}
  \caption{A magnified view of results near the lower end of the first
    order transition, including data for various time extents.  Note
    that there are two points close together for $N_T=8$, only one of
    which is shown.  The lines are drawn purely to guide the eye.  The
    dashed lines represent finite temperature effects for various
    $N_T$.  The thick lines are first order; the separation of the
    $N_T=2$ line is presumably an additional effect of the small
    lattice size.}
  \label{fig:closeup}
\end{figure}

Likewise, the complete results are summarised diagrammatically, the
overall picture in figure~\ref{fig:newres}, which does not show the
data points for clarity, and an expansion of the area near the lower
end of the first order transition including the data in
figure~\ref{fig:closeup}.

\begin{table}[tb]
\begin{center}
\begin{tabular}{l|rrrrrr}
\renewcommand{\arraystretch}{1.2}
$\beta_A$ & \multicolumn{6}{c}{$\beta_F$} \\
\hline\hline
 & \multicolumn{2}{c}{$N_T=2$} & \multicolumn{2}{c}{$N_T=4$} 
 & \multicolumn{2}{c}{$N_T=8$}\\
 & \multicolumn{1}{c}{$N_S=8$} & \multicolumn{1}{c}{$N_S=10$} & 
 \multicolumn{1}{c}{$N_S=8$} & \multicolumn{1}{c}{$N_S=10$} &
 \multicolumn{1}{c}{$N_S=16$} & \multicolumn{1}{c}{$N_S=18$}\\
\hline
1.0 & 1.654(9) & & 1.98(1) & 1.99(1) \\
1.7 & 1.365(6) & & 1.572(5) & 1.572(5) \\
2.2 & 1.166(2) & & 1.285(2) & 1.286(1) \\
2.4 & 1.090(1) & 1.090(1) & & & 1.190(2) & 1.189(2) \\
2.5 & 1.053(1) & 1.052(1) & & \\
\hline\hline
& \multicolumn{6}{c}{$\mathcal{O}(a)$ improved action} \\
& \multicolumn{1}{c}{$8^3\times4$} & \multicolumn{1}{c}{$10^3\times4$}\\
\hline
2.4 & 0.827(2) & 0.827(1) \\
\hline\hline
\end{tabular}
\end{center}
\caption{Results, using reweighting, for positions of phase transition
 for transitions with no clear metastability signal.}
\label{tab:ptpos}
\end{table}

\begin{table}[tb]
\begin{center}
\begin{tabular}{l|rrrrrr}
\renewcommand{\arraystretch}{1.2}
 $\beta_A$ & \multicolumn{6}{c}{$\chi_{P_t}$} \\
\hline\hline
 & \multicolumn{2}{c}{$N_T=2$} & \multicolumn{2}{c}{$N_T=4$} 
 & \multicolumn{2}{c}{$N_T=8$}\\
 & \multicolumn{1}{c}{$N_S=8$} & \multicolumn{1}{c}{$N_S=10$} & 
 \multicolumn{1}{c}{$N_S=8$} & \multicolumn{1}{c}{$N_S=10$} &
 \multicolumn{1}{c}{$N_S=16$} & \multicolumn{1}{c}{$N_S=18$} \\
\hline
1.0 & 0.0098(2) & & 0.0055(1) & 0.0043(1) \\
 & & & \multicolumn{2}{c}{(1.53(5))} \\
1.7 & 0.0119(3) & & 0.0065(1) & 0.0051(1) \\
 & & & \multicolumn{2}{c}{(1.53(4))} \\
2.2 & 0.0172(4) & & 0.0109(2) & 0.0087(2) \\
 & & & \multicolumn{2}{c}{(1.56(5))} \\
2.4 & 0.0251(5) & 0.0200(4) & & & $9.6(3)\times10^{-4}$ &
$8.1(4)\times10^{-4}$ \\ 
 & \multicolumn{2}{c}{(1.56(4))} & & & \multicolumn{2}{c}{(1.20(7))} \\
2.5 & 0.0327(5) & 0.0308(6) & & \\
 & \multicolumn{2}{c}{(1.84(5))} \\
\hline\hline
& \multicolumn{6}{c}{$\mathcal{O}(a)$ improved action} \\
& \multicolumn{1}{c}{$8^3\times4$} & \multicolumn{1}{c}{$10^3\times4$}\\
\hline
2.4 & 0.0108(4) & 0.0086(3) \\
 & \multicolumn{2}{c}{(1.56(8))} \\
\hline\hline
\end{tabular}
\end{center}
\caption{Results, using reweighting, for the temporal Polyakov loop
  susceptibility for transitions with no clear metastability signal.
  The numbers in parentheses show the ratio defined in equation
  \protect\ref{eqn:ratio}, expected to be 1.55 for a second order and
  1.95 for a first order transition for the spatial sizes used at
  $N_T=2$ and 4; for those at $N_T=8$ the numbers are 1.15 and 1.42
  respectively.}
\label{tab:ptsus}
\end{table}

\begin{table}[tb]
\begin{center}
\begin{tabular}{l|rrrrr}
\renewcommand{\arraystretch}{1.2}
$\beta_A$ & \multicolumn{5}{c}{$\beta_F$} \\
\hline\hline
 & \multicolumn{1}{c}{$8^3\times2$} & \multicolumn{1}{c}{$8^3\times4$} &
 \multicolumn{1}{c}{$12^3\times6$} & \multicolumn{1}{c}{$16^3\times8$}
 & \multicolumn{1}{c}{$12^4$} \\
\hline
2.4 & & 1.181(2) & 1.186(2) & 1.185(2) & 1.185(1) \\
2.5 & & 1.133(2) & 1.138(2) & \\
3.0 & 0.866(2) & 0.919(2) & 0.917(2) & & 0.917(1) \\
3.5 & 0.691(2) & 0.739(2) & 0.737(2) \\
\hline\hline
& \multicolumn{4}{c}{$\mathcal{O}(a)$ improved action} \\
& \multicolumn{1}{c}{$8^3\times4$} \\
\hline
2.5 & 0.79(1) \\
\hline\hline
\end{tabular}
\end{center}
\caption{Results for positions of phase transition
 showing a strong metastability signal.}
\label{tab:ptmeta}
\end{table}


\subsection{Location and nature of phase transitions}
Our initial goal was naturally to find the positions of the phase
transitions in the fundamental/Villain plane and their natures.
Some exploratory work was done on $N_S=6$ lattices, but for the
results quoted we have used $N_S=8$ as the smallest spatial size.

We have located the phase transition at a range of $(\beta_F,\beta_V)$
for both $N_T=2$ and 4.  The results are shown in
tables~\ref{tab:ptpos} to~\ref{tab:ptmeta}.
The first major point is that we confirm the results of \cite{ggm,mg}
that there is a change of nature in the phase transition.  Indeed, the
change is so clear that it necessitates a change of our methods of
analysis.


For small $\beta_V$, we encounter no problems and the analysis is that
of the standard finite temperature SU(2) transition.  The values
quoted for the phase transition are the positions of the peak in the
susceptibility $\chi_{P_t}$.  We concentrate on $N_T=4$ for detailed
verification of the order of the transition and have performed
simulations with $N_S=8$ and 10 to look at the scaling of the
peak in the susceptibility $\chi_{P_t}$.

For larger $\beta_V$, we indeed find that the transition has become
first order with clear two-state signals.  The transition becomes
stronger as $\beta_V$ is increased, with the time to flip between the
two states increasing from of the order of a thousand at $\beta_V=2.4$
to many tens of thousands of our (compound) sweeps.  Sufficiently
large statistics for looking in detail at the critical exponents over
a range of couplings are therefore beyond the scope of this paper, and
we have relied simply on the observation of a clear metastability
signal for determining the nature of these transitions.

The ratio defined in equation~\ref{eqn:ratio} is also shown in
table~\ref{tab:ptsus}.  From it and the onset of metastability, we can
deduce that the $N_T=2$ lattices become first order around
$\beta_V=2.5$ --- in fact, the value at this point suggests that the
lattice is in the tricritical region --- while for $N_T=4$ it is at a
slightly lower $\beta_V$, between 2.2 and 2.4.  This puzzling trend,
noted by us previously~\cite{me} and also by reference~\cite{gm}, will be
commented on below.

The values of the spatial Polyakov loop $P_x$ have been used as a check
for the presence of finite temperature behaviour; we observe in all
cases where $N_T>N_S$ that the $P_t$ symmetry breaking occurs
first for smaller $\beta_F$.  Caution is necessary, however; as we
shall see, the presence of a signal for finite temperature behaviour
does not necessarily mean we have identified the physical transition.

\subsection{Position of end point}
We have tried to estimate the position of the end point of the first
order effects for the $8^3\times4$ lattice. We find some quantity
which jumps across the transition: we have chosen the average
plaquette $W_\Box$ as it is easy to calculate.  The errors shown are just those
corresponding to our ignorance of the exact position of the phase
transition, deduced from reweighting the data to the central estimate
of the transition and to one standard error away.  These are quite
large, since in this region the average plaquette is changing
rapidly, which (to anticipate our later analysis) is presumably the
source of the first order effects.

We then fit to the formula expected for tricritical behaviour~\cite{ejnz}:
\begin{equation}
  \Delta W_\Box(\beta_V) = A\left(\beta_V - \beta_{V(\mathrm{tricrit})}\right)
^{\beta^u}
\label{eqn:tricrit}
\end{equation}
for the coefficients $A$, $\beta_{V(\mathrm{tricrit})}$ and $\beta_u$.
The result is shown in table~\ref{tab:tricrit}.  Our estimate for the
end point is therefore $\beta_{V(\mathrm{tricrit})} = 2.22(8)$.  In
fact, the results in table~\ref{tab:ptsus} tend to exclude the lower
part of the error range.  Also it is likely, given our comments below,
that there are significant finite size effects in this value.

\begin{table}[tb]
\begin{center}
\begin{tabular}{rl|rl}
\noalign{\renewcommand{\arraystretch}{1.1}}
\multicolumn{2}{c|}{Input values} & \multicolumn{2}{c}{Result} \\
$\beta_V$ & \multicolumn{1}{c|}{$\Delta W_\Box$} & \multicolumn{1}{c}{Coeff.}
& \multicolumn{1}{c}{Value} \\
\hline\hline
2.4 & 0.096(5) & $A$ & 0.365(15) \\
2.5 & 0.137(5) & $\beta_{V(\mathrm{TCP})}$ & 2.22(8) \\
2.6 & 0.169(9) & $\beta_u$ & 0.78(16) \\
3.0 & 0.301(3) & $\chi^2$ & 0.19 (1 d.o.f.) \\
\hline\hline
\end{tabular}
\end{center}
\caption{Results of the search for the tricritical end point on the
  $8^3\times4$ lattice.  The coefficients are defined in
  equation~\protect\ref{eqn:tricrit}.}
\label{tab:tricrit}
\end{table}


\subsection{Behaviour of first order transition}
We located the position of the transition initially by
performing simulations at fixed $\beta_V$ and varying $\beta_F$; we
did not see any sign of a second order, finite temperature
phase transition separate from the first order lines.  In fact, we
would not expect this, since examination of the temporal Polaykov loop
in the two different phases indicates that the low-$\beta_F$ phase is
confined while the high-$\beta_F$ phase is unconfined.  Again, the
difference between the phases increases with $\beta_V$.  We find no
such change in nature for spatial Polyakov loops.  It should also be
noted that all the quantities we have measured show this two state
signal; the behaviour is quite different to the usual finite
temperature transition where, on finite lattices, observables remain
continuous.

The next major question is whether the first order transition does
indeed have a finite temperature nature --- or, at the least, can be
connected with the finite size of the time direction of the lattice.
We need to verify the altered positions of the phase transition for
$N_T=2,$ 4 and preferably larger lattices.  Because the different
phases are stable for so long near the phase transition, this is
difficult by the usual methods.

We have therefore used the mixed start procedure to help locate the
transition~\cite{ejnz}.  In this method, one welds together a lattice
from parts in two phases.  If the transition between the phases is
sufficiently strongly first order --- as it appears to be in our case
--- this creates a pair of interfaces between the phases.  When this
lattice is updated further, the interface persists for some time (up
to a few hundred sweeps in our simulations), but eventually the more
stable phase is expected to predominate and the interface will
disappear.

There is a question as to how hard one needs to work to ensure
equality of probability for the two phases at the phase transition: in
the more sophisticated versions of the procedure~\cite{ejnz}, both
phases are prepared properly then joined together; the interface
between the two phases is then smoothed by updating the disordered
(our low temperature) phase only.  Reference~\cite{ejnz} found a few dozen
sweeps were necessary.  In our case, we find that only two or so of
these smoothing sweeps are required to achieve the objective, namely
that the average plaquette is roughly half way between the value in
the two pure phases.  In progressively simpler versions, the smoothing
does not take place, or the ordered phase is not initially
equilibrated and the links (and Z(2) variables in our case) set to
unity.  We have looked at all three versions as an attempt to
understand the systematic errors involved.

It seems that the simpler versions slightly underestimate the value of
$\beta_F$ for the transition, in other words favour the high
temperature phase.  The simplest of all (where the ordered phase is
inserted at infinite temperature) is lower than the other by a few
parts in ten thousand.  On the other hand, the smoothed version, even
with the few steps we have used, causes an even larger underestimate:
as much as 0.007 in the case of the transition for $8^3\times4$ at
$\beta_V=3.0$.  We therefore give our results from the other versions
to three decimal places, though we have calculated the next figure.
This is the source of the phase transition data in
table~\ref{tab:ptmeta} for all the first order transitions.  It should
be noted, therefore, that the values are likely to be underestimates,
though we do not know enough to be able to claim them as lower limits.

\subsection{Bulk or finite temperature?}
The results show an unambiguous separation between the $N_T=2$ and 4
results: what is more, this separation is maintained up to large
$\beta_V$, in fact as far as we have followed it.  The question arises
as to whether this is really a true finite temperature effect, or a
remnant thereof, or simply a finite size effect --- in other words, it
is possible that beyond a certain lattice size the first order effects
remain fixed.  We have therefore performed simulations for $N_T=6$
lattices in the region of the first order effects.

These $N_T=6$ results show a transition close to the $N_T=4$ results:
much more so than one would expect from a naive extrapolation between
the $N_T=2$ and 4 results.  Further, at larger $\beta_V$ the
transition point is the same within the errors of our method.  Thus it
seems quite likely that the whole of the first order line is
essentially bulk in nature, but with finite size, rather than finite
temperature, effects, and that these effects are largest in the region
near the first order end point, where presumably complex behaviour is
involved.  By `finite size' effects, we here mean numbers which reach a
plateau as the spatial lattice size is increased, as distinct from
the scaling behaviour seen as the temporal lattice size changes (which
is in some sense a finite size effect as well).

We have also simulated a $12^4$ lattice at $\beta_V=3.0$ and at
$\beta_V=2.2$. On these lattices the interface between the hot and
cold phases survives for many hundreds of sweeps, so the accuracy is
increased, although the computational effort is greater to achieve it.
We find a clear first order transition on this lattice in the region
$\beta_c=0.917(1)$ --- extremely close to that for $N_T=4$, in fact
the same within errors.  This tends to confirm the suggestion that the
change between $N_T=2$ and 4 is a finite size effect which simply
disappears for larger volumes, leaving the usual bulk transition.

This allows us to touch on the suggestion recently made by Gavai and
Mathur~\cite{gm}.  They suggest that the first order part of the line
may be a true finite temperature effect.  Further, they note the
effect (seen by us too) that the first order transition starts at
lower SO(3) coupling for $N_T=4$ than 2, and conjecture that this
trend may continue --- with implications for fundamental SU(2) at
large $N_T$.  In other words, they suggest the possibility that on
sufficiently large lattices the first order effect could be the true
one.

Our results would tend to suggest on the contrary that the $N_T=2$
case is actually anomalous, and that at larger $N_T$ the first order
transitions behave like those at $N_T=4$ and 6, so that the
tricritical point does not come to lower $\beta_V$.

In an effort to see if we can find a point at which the supposed bulk
and finite temperature effects have separated, we have also looked for
two-state behaviour on the $12^4$ lattice at $\beta_V=2.2$ and $2.4$
--- respectively slightly below and above the tricritical point of the
line for $N_T=4$.  The usual second order deconfinement transition on
this lattice would be too vague to be of use, however given the sharp
nature of the first order effects we would expect to be able to see
them if they were present, even incipiently.  We can see a two-state
signal in the plaquette value in the second case, though not in the
first.

More importantly, we do not see any sign of a finite temperature
signal even at $\beta_V=2.4$, in that the Polyakov loop retains its
unbroken symmetry on both sides of the transition.  Here, the two
state signal appears around $\beta_F=1.185$.  It should be noted that
our runs here are not high statistics: simply runs of a few thousand
sweeps from a cold and a hot start to look for the two state signal.
Nonetheless, even if the two state signal were to disappear with
sufficiently long runs, any incipient sign of the effect at exactly
the couplings where one sees the first order effects on smaller
lattices is surely not to be taken lightly.

\begin{figure}[tb]
\begin{center}
\psfig{file=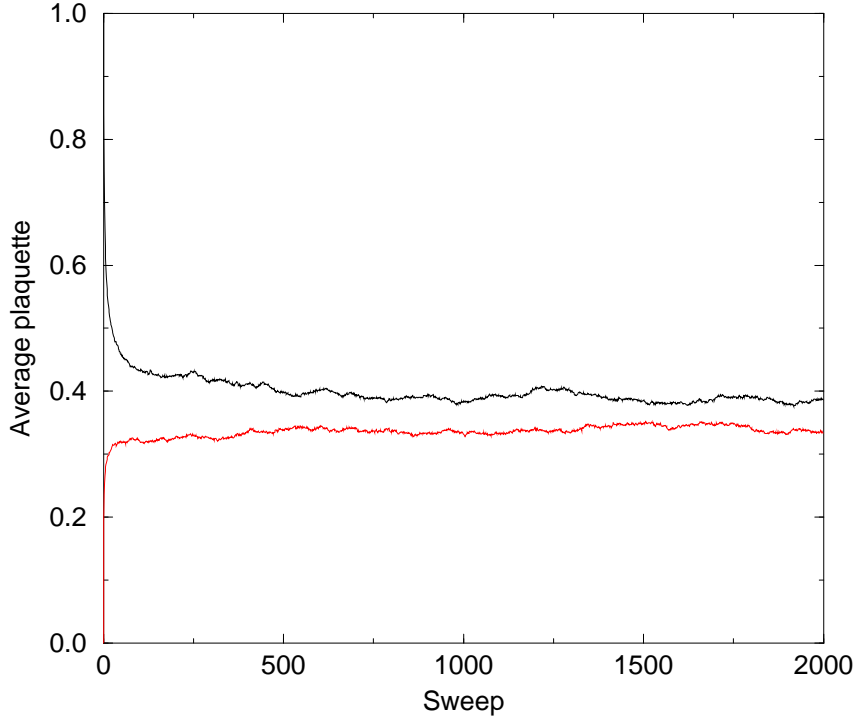,width=5in}
\end{center}
\caption{Metastability on the $16^3\times8$ lattices for
  $\beta_V=2.4$, $\beta_F=1.1852$.  This is near, but not necessarily
  right on top of, the first order transition.}
\label{fig:meta16_8}
\end{figure}

\begin{figure}[tb]
\begin{center}
\psfig{file=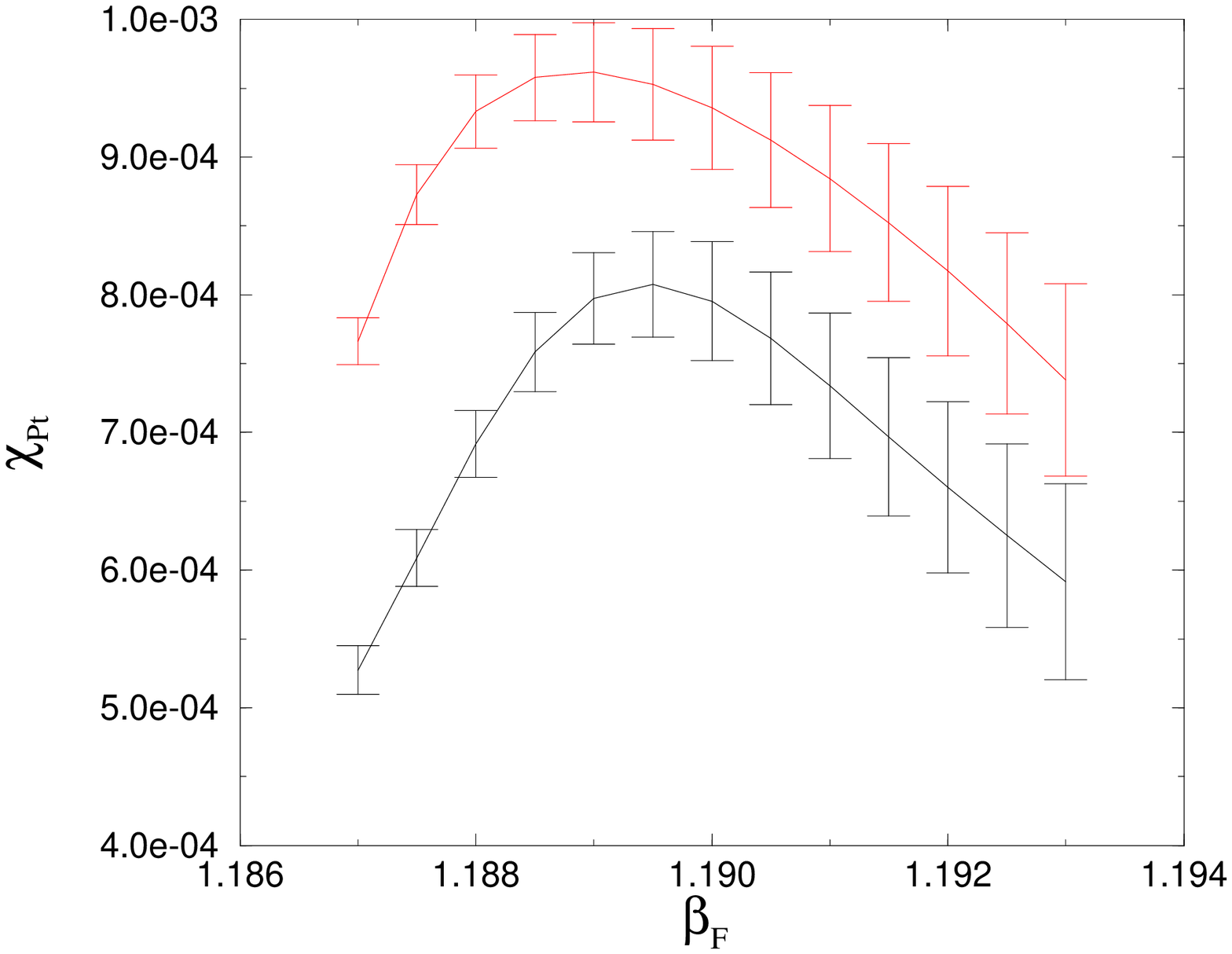,width=5in}
\end{center}
\caption{Peak in the temporal Polyakov loop susceptibility
  $\chi_{P_t}$ on the $16^3\times8$ (upper) and $18^3\times8$ lattices
  at $\beta_V=2.4$.}
\label{fig:ft8}
\end{figure}

These results strongly suggest it would be worthwhile to do a higher
statistics run on a larger lattice in this region.  We have therefore
simulated $16^3\times8$ and $18^3\times8$ lattices at $\beta_V=2.4$
just to the right of the first order effects.  The latter can clearly
be seen in figure~\ref{fig:meta16_8}; this may not be exactly at the
phase transition, due to the difficulties locating it on a lattice of
this size, but is close to it.  However, we can still pick out a
\textsl{separate} peak in the $\chi_{P_t}$ susceptibility at
$\beta_F=1.190(2)$, shown in figure~\ref{fig:ft8}, where our
simulations show no sign of metastability and are indeed all on the
high-$\beta_F$ side of the first order transition.  The range of
reweighting in the figure is limited to that which statistics seem to
allow: at least 1\% of the data corresponds to an average action whose
$\beta_F$ is more remote from the simulated value than the reweighting
point.  From table~\ref{tab:ptsus}, the ratio of
susceptibilities of equation~\ref{eqn:ratio}  is $1.20(7)$, compared
with the expectation 1.15 for a second order and 1.42 for a first
order transition.  This confirms that the transition is second order.

We consequently feel entitled to claim that the second order, finite
temperature effects separate out from the first order, bulk effects
near the lower end of the latter for time sizes $N_T\gtrsim8$, and
that this is the resolution of the universality problem.

Why should we see the apparent finite temperature effects, if the
transition is really bulk?  The following point may be relevant: there
is an interesting parallel between these effects and those seen in a
version of the compact lattice U(1) gauge theory, where one
forms a similarly modified action,
\begin{equation}
  S = -\sum_\Box(\beta\cos\theta_\Box + \gamma\cos2\theta_\Box)
\end{equation}
where $\theta_\Box$ is the plaquette angle~\cite{bhanot}.  Here again,
there is a phase transition --- although definitely bulk in the U(1) theory
--- which was found to change from second to first order as the
coupling with the extra symmetry, here $\gamma$, was increased.  In
reference~\cite{ejnz}, an investigation was made of this region, and the
tricritical point where the behaviour changed located.  Our point is
that near the tricritical region significant finite size effects were
seen, even on lattices of size $12^4$ and $14^4$.  Applied to
table~\ref{tab:ptmeta}, this fits in with our interpretation
of the results.



\section{Mixing of first and second order effects}
The evidence is that the finite temperature and the bulk transition do
separate out for sufficiently large lattices.  If one is interested
only in universality, and is prepared to dismiss the bulk effects as
an uninteresting lattice artefact, one can stop worrying at this
point.  We, however, shall discuss the nature of this further.

Heller has found\cite{heller} for the corresponding
fundamental/adjoint plane in SU(3) that there is also a clear
separation between the bulk effects and the deconfinement transition,
but that it occurs for rather smaller lattices; it seems that in this
case the bulk effects are not so strong. Possibly the difference lies
in topology of the gauge manifold.  Problems in the continuum limit of
certain gauge groups with a non-trivial topology like that of the
SO(3) case here have been raised~\cite{hh}.  The behaviour in the
representations of SU(3) has not yet been addressed, but it may be
that the effects there are sufficiently different to explain the
distinct results.

In explaining the mixing of the effects for lattices with
$N_T\lesssim8$, it is perhaps easiest to think of a more
basic picture of first order transitions, with no specifically
field-theoretic features.  The idea of physics being `hidden', as our
second order transition apparently is by the bulk effects, is quite
familiar from elementary thermodynamics.  Consider the water/steam
transition: metastability here is the property that allows one to
trace at least partly the hidden parts of the surface, as in
superheating or supercooling; a critical end point represents the
straightening out of the surface so that no such features occur.

We suggest thinking of the known bulk transitions in the SU(2)/SO(3)
model in a similar way.  For example, in increasing $\beta_V$ from one
side of the Z(2) monopole transition to the other, we are passing from
one range of physics in the gauge theory to another.  Without this
first order effect, the gauge dynamics would vary smoothly from strong
to weak coupling, so we are skipping over some inaccessible region ---
which quite likely includes, for example, whatever finite temperature
crossover is taking place in SO(3).

Presumably the conjunction of the two separate bulk transitions makes
this more complicated.  It would not be surprising if it had long
range effects; indeed, it has been suggested for a long time that the
famous peak in the specific heat of the plaquette of SU(2) ---
equivalently, the sharp crossover in the average action --- is a
long-range remnant of what was then thought to be a bulk end point.
This is reinforced by figure~\ref{fig:newres}: it is clear that even
where the deconfinement transition is visible as a second order line,
the lines for $N_T=2$ and 4 are converging, contrary to the naive
expectation that they are parallel at some effective beta given by
equation~\ref{eqn:beff}.

\begin{figure}[tb]
\begin{center}
\psfig{file=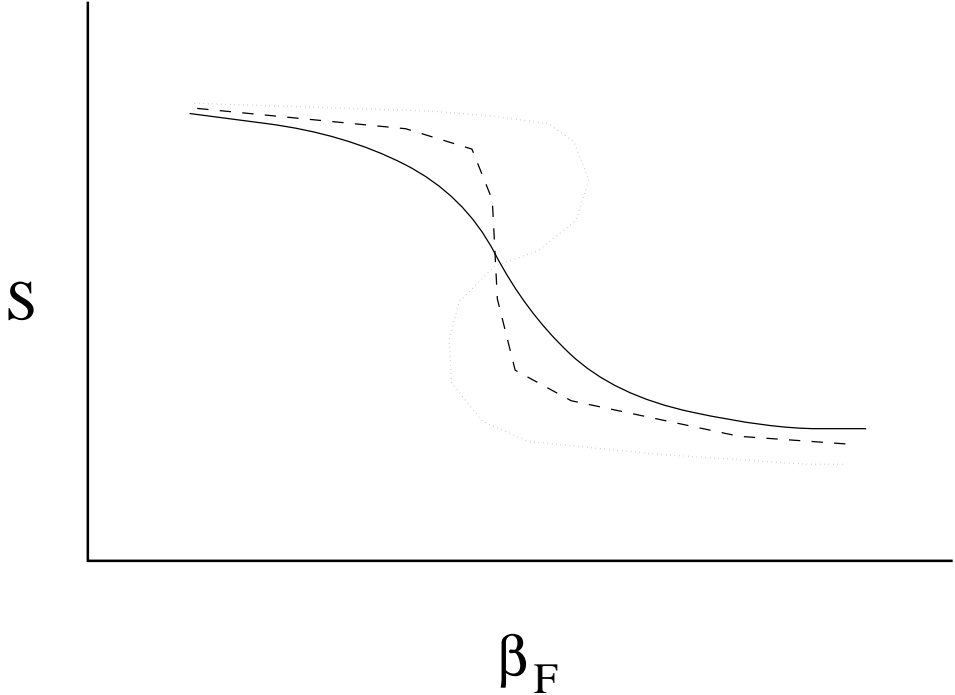,width=5in}
\end{center}
\caption{Freehand sketch of the presumed behaviour of the action in
  the crossover region of the average action below (full), at (dashed)
  and above (dotted) the end point of the first order transition.  The
  instability in the last case causes the discontinuous behaviour.}
\label{fig:endpoint}
\end{figure}

Figure~\ref{fig:endpoint} shows what we presume to be the behaviour of
the action below, on, and above the first order endpoint.  In the
third case (dotted line), the second order transition (along, of
course, with all other physics in the range of couplings) is hidden on
what one may think of as the `superheated' and `supercooled' branches
of the line, which are not seen because the instability in the action
causes it to flip from the high to the low side.  The tricritical
point (dashed line) is like the end point of the water/steam phase
transition in that the hidden regions disappear and the crossover from
one phase to another becomes smooth.  Our case is more complicated
because part of this smooth physics is the physical second order
transition.

This agrees with the observation that the first order transition
becomes stronger with increasing $\beta_V$ and that the discontinuity
in the average plaquette increases (as shown by our tricritical fit
above).  The dependence on the temporal extent of the lattice
indicates the sensitivity to the changes taking place, which may not
be finite temperature effects.  Then the deconfining transition itself
is invisible in the $(\beta_F,\beta_V)$ plane for sufficiently small
$N_T$.

The conclusions are supported by the other calculations of the monopole
$\langle M\rangle$ and charge $\langle E\rangle$, defined in
equation~\ref{eqn:monopole} and their effective values deduced from
the plaquette $\langle\bar M\rangle$ and $\langle\bar E\rangle$
defined in equation~\ref{eqn:effmonopole}.  The behaviour of these is
as given in reference~\cite{chs}; they too are discontinuous at the
first order lines, but show in any case a fairly sharp crossover from
near 1 at low $\beta_F$ to near 0 at high values.  We have nothing to
add in detail to the picture shown in figure~5 of
reference~\cite{chs}, which shows a larger range of $\beta_F$ (there
called $\delta$) than we have used.  The region where this occurs
appears to be the same as where the plaquette action shows a steep
change.  We consider this to be evidence in favour of the traditional
picture of the first order effects: that they are an instability
caused by the increasingly sharp crossover in the action.


\subsection{Extended action}
We supplement our evidence from the $12^4$ lattices, that the bulk and
finite temperature transitions can be seen separately at least at
$\beta_V=2.4$, with the following.

Given our contention that the first order effects do not represent
continuum physics, one might wonder whether an improved action such as
Symanzik's tree-level improved action~\cite{symanzik} would change the
picture, in that one is brought closer to the continuum for a given
lattice size by explicitly eliminating terms in the expansion of the
lattice action up to the next even power of $a$, $\mathcal{O}(a^6)$.
The action with this first order improvement is defined by:
\begin{equation}
  S_{Fi} = - \frac{\beta_{Fi}}{2}\left(\frac{5}{3}\sum_\Box \Tr_F U_\Box
- \frac{1}{2}\sum_{[1\times2]} \Tr_F U_{[1\times2]}\right)
\label{eqn:improved}
\end{equation}
in which the usual plaquette term has been supplemented by a sum over
all rotations of all $1\times2$ rectangles on the lattice and we use
the subscript $Fi$ to denote an improvement in the fundamental part of
the action.  Such actions have been shown to have problems when
looking at properties derived from two-point functions, presumably
related to their lack of positivity~\cite{mt}; also, we do not know if
the perturbative coefficients are close to the optimum for
non-perturbative simulations.  However, there should be no problem in
simply looking for some sort of improvement in a phase transition with
a naive perturbative improvement.

\begin{figure}[tb]
\begin{center}
\psfig{file=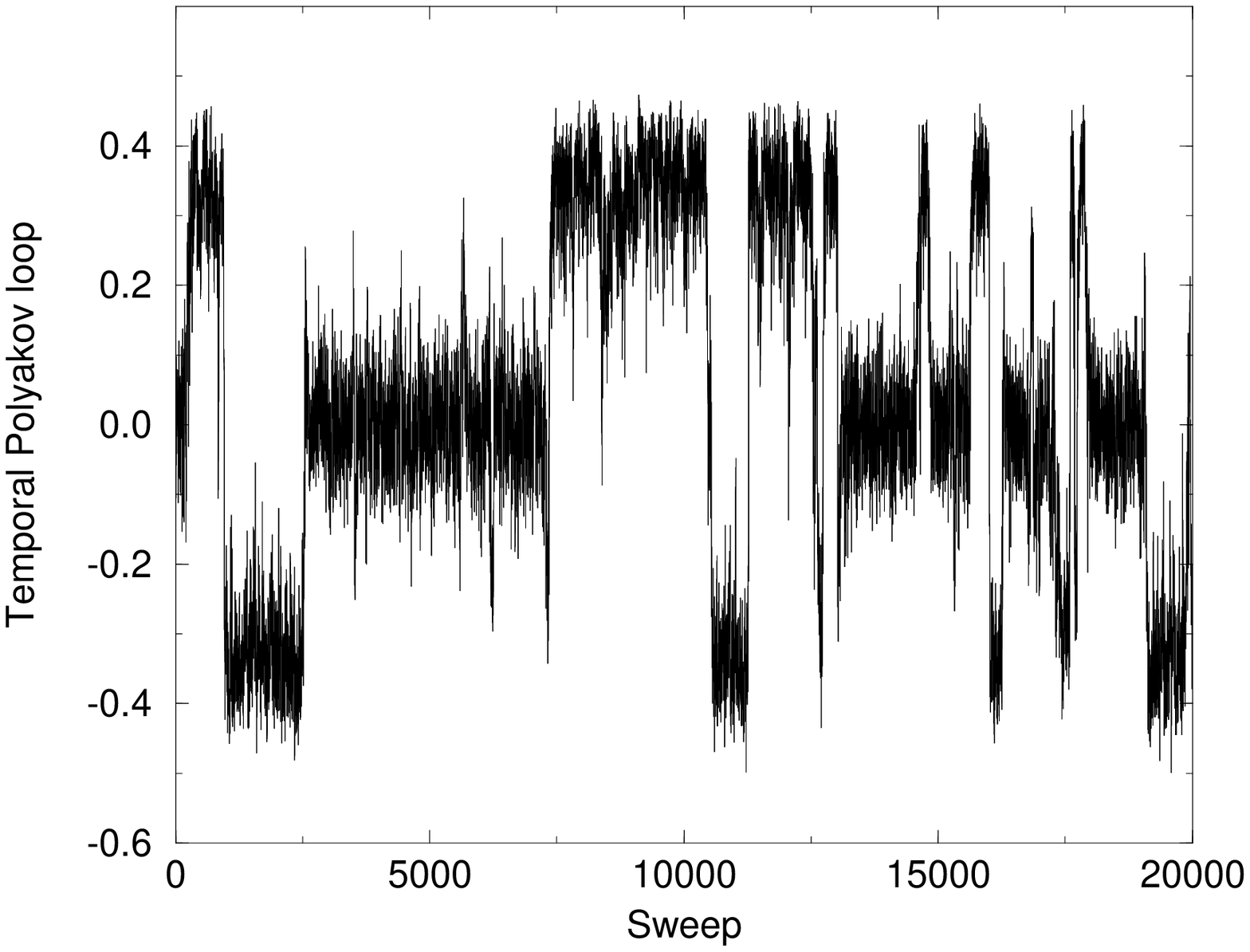,width=5in}
\end{center}
\caption{Metastability for the unimproved $8^3\times4$ lattice with
  $\beta_F=1.181$, $\beta_V=2.4$, showing the average Polyakov loop in
  the finite temperature direction.}
\label{fig:meta8_4}
\end{figure}

We have simulated using the action in equation~\ref{eqn:improved} for the
fundamental coupling only.  This will allow us to use the unchanged
$\beta_V$ axis as a yardstick for how well the improved action is
performing.  We again pick $N_T=4$ for $N_S=8$ and 10 at $\beta_V=2.4$.
To demonstrate that in the unimproved case there is metastability
suggestive of a first order transition, we show the behaviour of the
temporal Polyakov loop on a lattice of the same size at
$\beta_F=1.181$ in figure~\ref{fig:meta8_4}: it is clear there is a
two state signal, with one phase confined, the other deconfined.

With the improved action we find no sign of first order effects.  We
instead find a smooth change of behaviour with a peak in $\chi_{P_t}$
at $\beta_{Fi}=0.827(2)$; the scaling behaviour clearly suggests a
second order nature (table~\ref{tab:ptsus}).
(Note that there is a rescaling of $\beta_F$ in the improved case;
this is not important so long as we can identify the corresponding
physical regions.)

There is thus a clear --- in fact, a qualitative --- difference of
behaviour between the unimproved and improved action at the same
$\beta_V$.  As we are using the same Villain part of the action, we
can be sure the difference is due to the improvement in the
fundamental part.  We hold this to be evidence for our contention that
the first order effects are artefacts irrelevant to the continuum
limit of the gauge theory.  There may still be some continuum limit at
the end of the first order transition, but we suggest it is not simply
related to that of the usual SU(2) theory.

We have also simulated at $\beta_V=2.5$: here the improved action too
shows the two state behaviour, with a transition in the vicinity of
$\beta_{Fi}\sim0.79$, so (as one would expect) the effect of the
first-order improvement is fairly small.


\section{Conclusions}

We have attempted to resolve the confusions found in pure SU(2)
lattice gauge theory with a mixed fundamental/SO(3) action.  We have
used the Halliday--Schwimmer action which allows efficient updating.
Our conclusions are summarised in figure~\ref{fig:newres}, where for
clarity the data points are not shown and should be read off from
tables~\ref{tab:ptpos} to~\ref{tab:ptmeta}, and in an expansion of
the area around the endpoint of the first order behaviour in
figure~\ref{fig:closeup}.

We confirm that for lattices with small extension $N_T$ the finite
temperature SU(2) transition runs into a region where first order
effects dominate.

On small lattices, between $N_T=2$ and $4$ sites in the finite
temperature direction, we see a shift in the position of the first
order effects to higher fundamental inverse coupling $\beta_F$, as
seen by references~\cite{ggm,mg,gm}.  For larger $N_T$, and at
slightly increased SO(3) coupling $\beta_V$, this effect decreases and
we suggest it is due to large finite \textsl{size} (in the sense that
they disappear on larger lattices) rather than \textsl{temperature}
effects.  This supports the traditional view that the first order
effects have a bulk nature and are unrelated to the true finite
temperature transition.

We have reinforced the suggestion that the first order effects are
not related to the continuum limit of SU(2) by showing that they are
shifted to larger $\beta_V$ by the use of an order-$a$ improved gauge
action.

The evidence we have presented suggests that the finite temperature
transition will separate out from the bulk effects, and has indeed
started to do so at the lower end of the bulk effects, around
$\beta_V=2.4$, for $N_T=8$.  In this case (in contrast to SU(3)) the
process is likely to be gradual: at $\beta_V=3.0$, for example, the
$12^4$ lattice again shows Polyakov loop symmetry breaking across the
first order transition, showing that the finite temperature effects
have been reabsorbed into the bulk ones.  There is therefore no reason
to doubt that eventually the former will emerge cleanly from the
latter for a lattice with a sufficiently large temporal extent.
Clearly, this would require huge lattices and statistics to sort out
quantitatively; eventually one has to worry about the intersection
with the Z(2) symmetry breaking transition at large $\beta_V$.  This
is probably out of reach at the moment.

Nonetheless, we suggest that the problems of universality in this
theory are essentially resolved, and that the true finite temperature
transition in the extended SU(2) plain remains second order, while the
first order effects present are bulk artefacts which are modified by
small lattice dimensions.

While this work was being finalised, a new preprint appeared further
disputing the nature of the first order line in the region containing
the endpoint, based on a finite size scaling analysis~\cite{rvg}.  To
resolve this would involve a more detailed investigation of the
tricritical region.


\section*{Acknowledgments}

I should like to thank Simon Hands for many discussions during the
work on this paper, Urs Heller for helpful comments and suggestions,
and Karl Jansen and Chuan Liu for an enlightening discussion.



\end{document}